\documentclass[aps,prl,reprint,amsmath,amssymb,superscriptaddress,nobibnotes,showpacs,showkeys,floatfix]{revtex4-1}

\usepackage{lipsum}

\usepackage[utf8]{inputenc}
\usepackage[T1]{fontenc}

\usepackage{graphicx}

\usepackage{dcolumn}

\usepackage{bm}

\usepackage{amsmath}

\usepackage{amssymb,amsthm}

\usepackage{bbm}

\usepackage{epsfig,color}

\usepackage[sans]{dsfont}

\usepackage{comment}

\usepackage{hyperref}

\usepackage{color}

\usepackage{tikz}

\usepackage{tikz-3dplot}

\usetikzlibrary{patterns}

\usepackage{ifthen}

\usepackage{pgfplots}

\usepackage{graphicx}

\graphicspath{ {images/} }

\definecolor{nblue}{rgb}{0.2,0.2,0.7}

\definecolor{ngreen}{rgb}{0.2,0.6,0.2}

\definecolor{nred}{rgb}{0.7,0.2,0.2}

\definecolor{nblack}{rgb}{0,0,0}

\newcommand{\tr}{\text{tr}}

\newcommand{\pguess}{p_\textrm{guess}}

  \theoremstyle{definition}

  \theoremstyle{plain}

\theoremstyle{plain}

\theoremstyle{plain}

  \theoremstyle{plain}

  \theoremstyle{plain}

  \providecommand{\conjecturename}{Conjecture}

  \providecommand{\definitionname}{Definition}

  \providecommand{\lemmaname}{Lemma}

\providecommand{\corollaryname}{Corollary}

\providecommand{\theoremname}{Theorem}

\providecommand{\propositionname}{Proposition}

\def\i{\mathrm{id}}

\def\g{\mathrm{guess}}

\def\E{\mathcal{E}}

\def\N{\mathcal{N}}

\def\T{\mathcal{T}}

\def\tr{\mbox{tr}}

\def\bea{\begin{eqnarray}}

\def\eea{\end{eqnarray}}

\begin{document}




\title { Measurement-Protected Quantum Key Distribution } 


\author{Spiros Kechrimparis} 
\email{Equally contributed}
\affiliation {School of Electrical Engineering, Korea Advanced Institute of Science and Technology (KAIST), 291 Daehak-ro Yuseong-gu, Daejeon 34141 Republic of Korea.}

\author{Heasin Ko} \email{Equally contributed}
\affiliation{Photonic \& Wireless Convergence Components Research Division, Electronics and Telecommunications Research Institute (ETRI), Daejeon 34129, Republic of Korea}

\author{Young-Ho Ko}
\affiliation{Photonic \& Wireless Convergence Components Research Division, Electronics and Telecommunications Research Institute (ETRI), Daejeon 34129, Republic of Korea}

\author{Kap-Joong Kim}
\affiliation{Photonic \& Wireless Convergence Components Research Division, Electronics and Telecommunications Research Institute (ETRI), Daejeon 34129, Republic of Korea}

\author{Byung-Seok Choi}
\affiliation{Photonic \& Wireless Convergence Components Research Division, Electronics and Telecommunications Research Institute (ETRI), Daejeon 34129, Republic of Korea}

\author{ Chahan M. Kropf}
\affiliation{ Istituto Nazionale di Fisica Nucleare, Sezione di Pavia, via Bassi 6, I-27100 Pavia, Italy }
\affiliation{Dipartimento di Matematica e Fisica and Interdisciplinary Laboratories for Advanced Materials Physics, Universit\`{a} Cattolica del Sacro Cuore, via Musei 41, I-25121 Brescia, Italy}

\author{Chun Ju Youn}
\email{ Corresponding author: cjyoun@etri.re.kr}
\affiliation{Photonic \& Wireless Convergence Components Research Division, Electronics and Telecommunications Research Institute (ETRI), Daejeon 34129, Republic of Korea}

\author{ Joonwoo Bae }
\email{ Corresponding author: joonwoo.bae@kaist.ac.kr}
\affiliation {School of Electrical Engineering, Korea Advanced Institute of Science and Technology (KAIST), 291 Daehak-ro Yuseong-gu, Daejeon 34141 Republic of Korea.}

\begin{abstract}

In the distribution of quantum states over a long distance, not only are quantum states corrupted by interactions with an environment but also a measurement setting should be re-aligned such that detection events can be ensured for the resulting states. In this work, we present measurement-protected quantum key distribution where a measurement is protected against the interactions quantum states experience during the transmission, without the verification of a channel. As a result, a receiver does not have to revise the measurement that has been prepared in a noiseless scenario since it would remain ever optimal. The measurement protection is achieved by applications of local unitary transformations before and after the transmission, that leads to a supermap transforming an arbitrary channel to a depolarization one. An experimental demonstration is presented with the polarization encoding on photonic qubits. It is shown that the security bounds for prepare-and-measure protocols can be improved, for instance, errors up to $20.7\%$ can be tolerated in the Bennett-Brassard 1984 protocol.
\end{abstract}

\maketitle


Recent advances in quantum communication technologies have successfully demonstrated the distribution of quantum states over a long distance, such as the quantum-enabled satellite \cite{liao2017, liao2018, bedington2017}, an airborne platform \cite{pugh2017} and submarine telecommunication fiber \cite{Wengerowsky6684}. Quantum key distribution (QKD) protocols, the art of establishing secret communication between honest parties at a distance, have been also demonstrated. Toward real-world applications, it is significant that the protocols can deal with potentially existing, uncontrollable, and unknown sources of noise on quantum states during the transmission, e.g., the weather conditions, atmospheric effects, and so on \cite{Crane:1977aa, Burton:2015aa, Haarig:2018aa, Vasylyev:2019aa}. It is crucial to maintain not too high quantum-bit-error-rates (QBERs), since the error-rate is a single parameter from which two honest parties can estimate the intervention of an eavesdropper and subsequently decide if they abort or proceed to secret key distillation. 


From a fundamental point of view, detection events depend on both states and measurements: not only are quantum states corrupted by a noisy channel, but also measurements should be realigned for optimal detection of events. An optimal measurement that minimizes the effect of a noisy channel is generally sought. In search of a measurement for optimal detection, it may be required to verify the resulting states, for instance, by identifying a noisy channel. 


From a practical point of view, however, the verification of resulting states after an unknown channel requires a significant amount of experimental resources. Moreover, sources of noise may change over time and also depend on location. This also implies that the verification may have to be repeated whenever different channels are introduced. The consequence is that it becomes practically an infeasible process to revise a measurement setting for noisy transmission of quantum states. With a high rate of errors, QKD protocols would fail to establish secure communication. 

It is therefore left to devise a quantum protocol that builds an optimal measurement for the states sent over an unknown channel. Notwithstanding the no-go theorems when states are subsequently unknown \cite{Wootters:1982aa, Dieks:1982aa, PhysRevA.60.R2626, De-Martini:2002aa, PhysRevLett.83.432}, it has been shown that an optimal measurement can be preserved when the states before a channel are known \cite{PhysRevA.99.062302}. This is achieved by manipulating unknown quantum channels by local operations and classical communication, known as the framework of a supermap \cite{chiribella2008}, such that a measurement in a noiseless scenario is kept ever optimal in a noisy environment \cite{spiros2019}.

We here propose a measurement-protected (MP) QKD protocol that aims to establish secure communication by keeping a single measurement setting ever optimal in the distribution of quantum states, without the necessity of verifying a channel connecting two honest parties. Thus, an optimal measurement setting prepared in a noiseless scenario can be re-used as an optimal one in a noisy environment. The measurement protection is achieved by including applications of a few local unitary transformations before and after the distribution of quantum states, by which an unknown channel is manipulated to preserve an optimal measurement. We show that in doing so, i) the distinguishability of quantum states can be enhanced, meaning that quantum states can be distributed over even longer distances, ii) a measurement prepared from the beginning would remain ever optimal, so that the verification of a channel is thus unnecessary, and iii) QBERs can be made lower, so that a higher error rate from an unknown channel can be tolerated for secret key distillation, e.g., up to $20.7\%$ in the Bennett-Brassard 1984 (BB84) protocol. We present proof-of-principle experimental demonstrations with photonic qubits. It is experimentally shown that by MP protocols, quantum states can be better distinguished over a distance and QBER can be made even lower. 

We begin by describing a prepare-and-measure (P\&M) protocol of two honest parties called Alice and Bob, with an example of the BB84 protocol \cite{bb84}. We suppose that Alice prepares quantum states and sends them to Bob. For instance, the four states
\bea
S_4 = \{ |0\rangle, |1\rangle, |+\rangle, |-\rangle \}  \label{eq:fourstate}
\eea
where $|\pm\rangle =  (|0\rangle \pm |1\rangle)/\sqrt{2} $, are exploited in the BB84 protocol. Bob applies a measurement with POVM elements $M_4 = \{ M_0, M_1, M_{+} , M_{-} \}$ where $M_{k} = | k \rangle \langle k|$ for $k=0,1,+,-$. The BB84 scenario in which Alice chooses one of the sets $\{ |0\rangle, |1\rangle \}$ or $\{ |+\rangle, |-\rangle \}$ is equivalent to that of minimum-error discrimination of the four states \cite{bae2013}. The measurement outcomes are sifted by public discussions between Alice and Bob, e.g., in the BB84 protocol, the bases used in the preparation and the measurement are announced. The classical post-processing proceeds to distilling secret bits with one-way information reconciliation and privacy amplification, see e.g., \cite{renner2008}.

P\&M protocols can be equivalently analyzed in the entanglement-based picture as follows \cite{bennett1992}. Instead of the preparation of quantum states, Alice firstly generates maximally entangled states $|\phi_1\rangle = (|00\rangle + |11\rangle)/\sqrt{2}$, then measures qubits in the first half, and sends the others to Bob through a channel. The analysis to obtain general security has shown that one can safely restrict to the case where two honest parties share symmetric states \cite{kraus2005, renner2005a, renner2008}. Then, a shared state on a single-copy level is given by a Bell-diagonal state $\rho_{AB} = (\mathrm{id} \otimes \N_{\bf p}) [| \phi_1\rangle \langle \phi_1 |]$ where a Pauli channel is denoted by
\bea
\N_{\bf{p} }[\rho] = p_0 \rho + p_{x} X\rho X + p_{y} Y\rho Y+ p_{z} Z \rho Z \label{eq:pauli}
\eea
with a probability vector ${\bf p} = [ p_0, p_x, p_y, p_z ]$ and Pauli matrices  $X$, $Y$, and $Z$. In particular, a depolarization channel is obtained when $p_x = p_y = p_z$. Since a state sent by Alice is noisy by a channel $\N_{\bf p}$, Bob has to update his measurement accordingly for optimal detection.

MP QKD protocols incorporate the measurement protection to the stage of distributing quantum states. This aims to keep Bob's measurement ever optimal for an ensemble of states even if the states are affected by an unknown channel. The main idea for the measurement protection is to manipulate a channel $\N_{\bf p}$ by a supermap $ \mathcal{\widetilde{S}}: \N_{\bf p}\mapsto \mathcal{\widetilde{S}} \N_{\bf p}$: for an ensemble $S$, a set of unitaries $ \mathcal{V} = \{V_j  \}$ is applied as follows,
\bea
\mathcal{\widetilde{S}} \N_{\bf p} [\rho] := \frac{1}{ |\mathcal{V} | } \sum_{V_j \in \mathcal{V} } V_{j}^{\dagger} \N_{\bf p} [ V_j \rho V_{j}^{\dagger}] V_j  \label{eq:ct}
\eea
for a state $\rho \in S$, such that the map $\mathcal{\widetilde{S}}  \N_{\bf p}$ corresponds to a depolarization channel,
\bea
D_{\eta} [\rho] = (1-\eta)\rho + \eta \frac{\mathbb{I} }{2},~\mathrm{where}~\eta = \frac{4}{3} (1-p_0). \label{eq:dep}
\eea
Consequently, regardless of what channels exist between Alice and Bob, it suffices to deal with depolarization noise in Eq. (\ref{eq:dep}), over which it has been shown that an optimal measurement for the ensemble $S$ is preserved \cite{PhysRevA.99.062302, spiros2019}. The verification of a channel has been circumvented. In practice, the protocol in Eq. (\ref{eq:ct}) can be realized by Alice randomly applying $V_j \in \mathcal{V} $ to a state and announcing to Bob her choice $j$, so that he can also apply $V_{j}^{\dagger} \in \mathcal{V}$ before his measurement.


The transformation of channels in Eqs. (\ref{eq:ct}) and (\ref{eq:dep}) for an arbitrary set of states can be achieved by a unitary $2$-design \cite{PhysRevA.80.012304}. In general, a unitary $2$-design has a cardinality $\mathcal{O} (d^4)$ for $d$-dimensional systems. For instance, for qubit channels there are $12$ unitaries that realize twirling, which are also conjectured to be minimal \cite{doi:10.1063/1.2716992}. By specifying an ensemble of states, the set of unitaries that achieve the transformation in Eqs. (\ref{eq:ct}) and (\ref{eq:dep}) can be further simplified. In the case of Pauli channels for qubit states, it holds that three unitaries which form a subset of the unitary $2$-design with the $12$ elements can generally construct the supermap in Eq. (\ref{eq:ct}). Note that the $12$ elements denoted by $\{ U_i \}_{i=1}^{12}$ are given by, 
\bea
&&U_{1} = \mathbb{I}, ~U_2 =  iX,~U_3= iY,~U_4 = iZ ~\label{eq:u2d}  \\
&& U_5 = \frac{1}{2} \left[ \begin{array}{ccc} 1-i & -1-i \\ 1-i  & 1+ i  \end{array} \right],~U_6 = \frac{1}{2} \left[ \begin{array}{ccc} 1+ i &  1-i \\ -1-i  & 1- i  \end{array}  \right] \nonumber \\ 
&& U_7 = \frac{1}{2} \left[ \begin{array}{ccc} 1 + i & -1 + i \\ 1 + i  & 1- i  \end{array} \right],~U_8 = \frac{1}{2} \left[ \begin{array}{ccc} 1- i &  1+ i \\ -1+ i  & 1+ i  \end{array}  \right] \nonumber \\ 
&& U_9 = \frac{1}{2} \left[ \begin{array}{ccc} -1 - i & -1 - i \\ 1 - i  & - 1 + i  \end{array} \right],~U_{10} = \frac{1}{2} \left[ \begin{array}{ccc} -1+  i &  1 -  i \\ -1 - i  & -1 - i  \end{array}  \right] \nonumber \\ 
&& U_{11} = \frac{1}{2} \left[ \begin{array}{ccc} -1 + i & -1 + i \\ 1 + i  & - 1 - i  \end{array} \right],~U_{12} = \frac{1}{2} \left[ \begin{array}{ccc} -1-  i &  1 +  i \\ -1 + i  & -1 + i  \end{array}  \right].  \nonumber 
\eea
Though not being unique the three elements can be chosen as $\{U, V, W\}$ with $U\in \{U_1, U_2, U_3, U_4 \}$, $V\in \{U_5,U_6,U_7,U_8 \} $, and $W\in \{ U_9,U_{10}, U_{11},U_{12}\}$. Thus, compared to standard P\&M QKD protocols, the measurement protection only requires applications of few unitary transformations before and after distributing quantum states, which we call quantum pre- and post-processing, respectively.


\begin{figure*}[t]
\begin{center}
\includegraphics[width=7.2in]{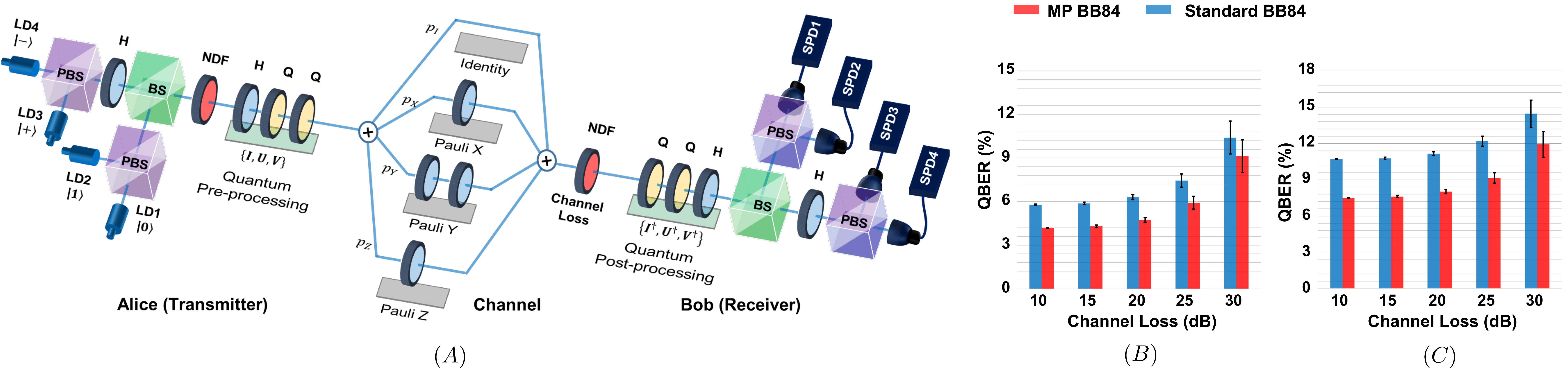}
\caption{ An experimental setup of the MP BB84 protocols is shown in $(A)$. States in the set $S_4$ are generated by polarizing beam splitters (PBS) and a beam splitter (BS). As a quantum pre-processing, local unitaries $\{\mathbbm{I}, U, V \}$ prepared by a set of half (HWP) and quarter (QWP) wave-plates are randomly applied. After a Pauli channel, see Eq. (\ref{eq:pauli}), a channel loss is simulated by a neutral density (ND) filter. The quantum post-processing is performed with the inverse of the unitary transformations. With the setup, QBERs of the BB84 protocol over the channel in Eq. (\ref{eq:ych}) are experimentally estimated in (B) and (C) for $p=0.05$ and $p=0.1$, respectively. The MP BB84 protocol has a lower value of QBER than the standard one. The channel loss due to  transmission over a long distance is simulated by ND filter from $10 dB$ to $30 dB$. QBERs can be suppressed by the measurement protection. } 
\label{setup}
\end{center}
\end{figure*}


An experimental setup to realize an MP BB84 protocol with polarization encodings on photonic qubits is shown in $(A)$ in Fig. \ref{setup}. In (B) and (C), it is experimentally demonstrated that a QBER can be made smaller by the measurement protection. The detailed information of the experiment is as follows. Four different polarization states in $S_4$, i.e., $|0\rangle = |H\rangle$, $|1\rangle = |V\rangle$, $|+\rangle = |D\rangle$ and $|-\rangle = |A\rangle$ where $H$ and $V$ ($D$ and $A$) are from the linear (diagonal) polarization, are generated with four vertical-cavity surface-emitting lasers. Photon pulses are attenuated with the mean photon number $0.1$ at a clock rate of $100$ MHz. Unitary transformations for the quantum pre- and post- processing are realized by a set of wave-plates. Pauli channels can be implemented by using half-wave plates. Then, the distribution of the states over a long-distance is simulated by ND filters, see $(A)$ in Fig. \ref{setup}. QBERs for the channel in Eq. (\ref{eq:ych}) are estimated. Photon detection is performed with a Silicon-Avalanche Photodiode based four-channel single photon detector where the dark count rate is lower than $500$ counts/s per channel and the detection efficiency is about $50 \%$. The loss during a measurement is measured about $8$ dB including optics insertion loss, coupling loss, sifting loss, and detection loss. 

\begin{figure}[t]
\begin{center}
\includegraphics[width=3.5in]{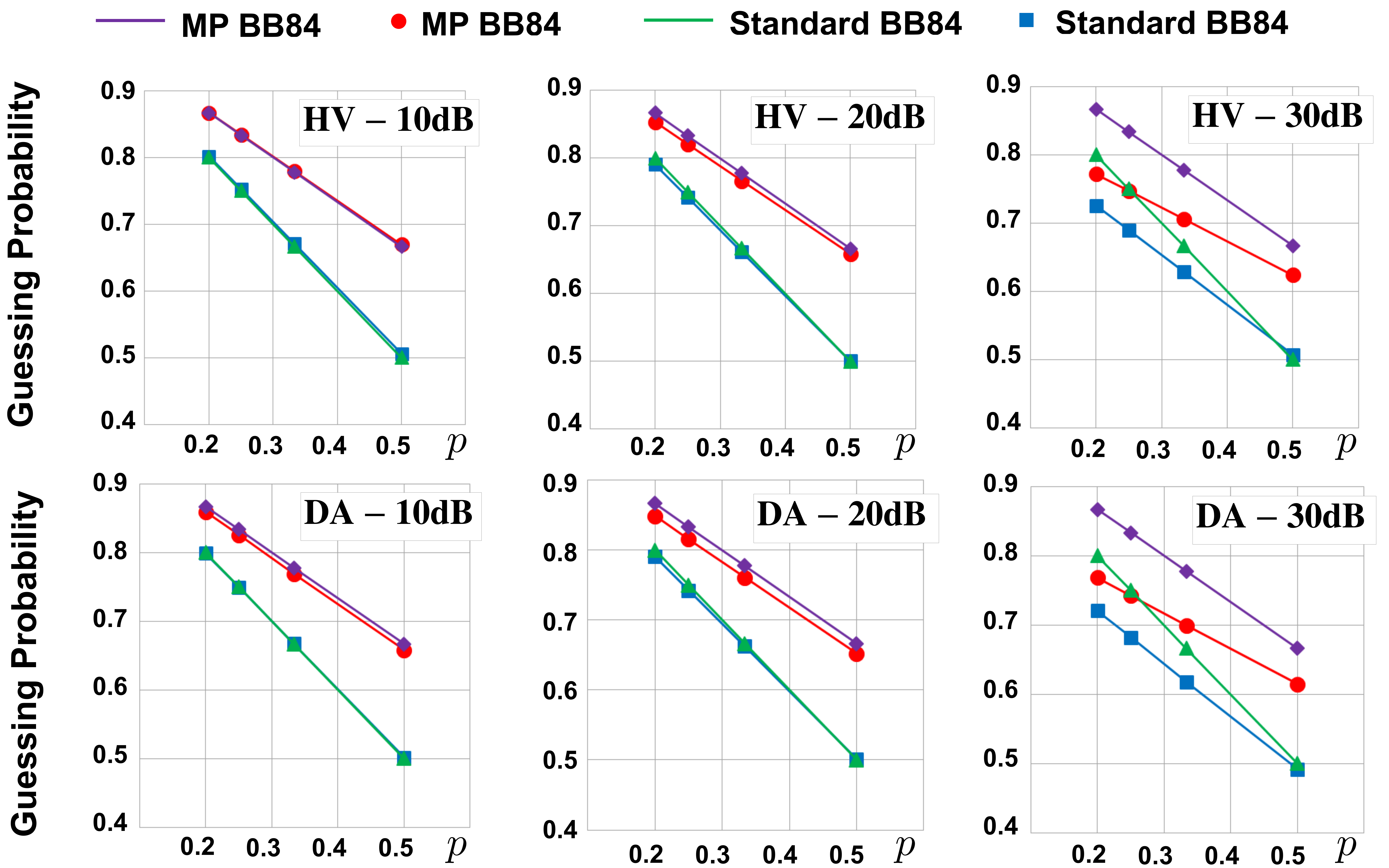}
\caption{ The guessing probabilities for ensembles of orthogonal states $\{|H\rangle, |V\rangle \}$ and $\{|D\rangle, |A\rangle \}$ over the channel in Eq. (\ref{eq:ych}) are shown. Solid lines are from Eq. (\ref{eq:s2g}) and dots are experimental data, see Fig. \ref{setup}. In all cases, two honest parties can achieve better distinguishability of quantum states over a quantum channel by using MP protocols. }
\label{gphv}
\end{center}
\end{figure}

Apart from applications to cryptographic protocols, the measurement protection can be also used to amplify distinguishability of quantum states over a noisy channel. As an example, a measurement $\{ |0\rangle \langle 0|, |1\rangle\langle 1| \}$ is optimal for a pair of distinguishable states $S_2 = \{|0\rangle,|1\rangle \} \subset S_4$. Suppose that the states are sent through a Pauli channel
\bea
\E[\rho] = (1-p)\rho + p Y\rho Y,~\mathrm{for}~ p\in[0,1/2]. \label{eq:ych}
\eea
Let $p_{\g}^{(\E)}$ denote the guessing probability with the channel $\E$, the highest success probability by minimizing the average error. From the Helstrom bound \cite{helstrom1976} (see reviews \cite{bergou2004, barnett2009, bae2015}), the guessing probabilities with and without the measurement protection, respectively, can be found as 
\bea
p_{\g}^{(\E)} = \frac{1}{2} + \frac{ 1-2p}{2} ~\mathrm{and}~ p_{\g}^{( \mathcal{\widetilde{S}}\E)} = \frac{1}{2} +  \frac{3-4p}{6}. ~~ \label{eq:s2g}
\eea
Clearly, it holds that $p_{\g}^{(\mathcal{\widetilde{S}}\E)} > p_{\g}^{(\E)}$. In particular, for the channel with $p=1/2$ the resulting states become identical and thus completely indistinguishable i.e. $p_{\g}^{(\E)} =1/2$. The measurement protection makes it possible to distinguish the states better than random, $p_{\g}^{ (\mathcal{\widetilde{S}}\E) } = 2/3$. In Fig. \ref{gphv}, an experimental demonstration is presented, which shows that distinguishability of quantum states can be amplified by an MP protocol. The experimental setup in Fig. \ref{setup} is exploited for pairs of orthogonal states $\{|0\rangle, |1\rangle \}$ and $\{|+\rangle, |-\rangle \}$ with linear and diagonal polarizations. 

For a pair of non-orthogonal states $S_{0+} = \{ |0\rangle, |+\rangle\}$ that can be used in the Bennett 1992 protocol \cite{bennett1992}, the guessing probability is given by $1/2 + 1/(2\sqrt{2})$ with the optimal measurement $M_{0+} = \{ |m_0\rangle \langle m_0| ,|m_{+}\rangle \langle m_{+}|  \}$ where $|m_0\rangle = \cos(\pi/8) |0\rangle -\sin (\pi/8) |1\rangle$ and $|m_+\rangle = \cos (3\pi/8) |0\rangle + \sin ( 3 \pi/8) |1\rangle$. For a Pauli channel in Eq. (\ref{eq:ych}), one can compute the guessing probabilities for cases with and without the measurement protection, respectively,
\bea
p_{\g}^{(\E)} = \frac{1}{2} + \frac{ 1-2p }{2\sqrt{2}}   ~\mathrm{and}~ p_{\g}^{ (\mathcal{\widetilde{S}}\E)} = \frac{1}{2} +  \frac{3-4p}{6\sqrt{2}}. ~~ \label{eq:s0+g}
\eea
Note that the guessing probability $p_{\g}^{ (\mathcal{\widetilde{S}}\E)}$ is obtained from the measurement $M_{0+}$ which has been used for the ensemble $S_{0+} = \{ |0\rangle, |+\rangle\}$. The other one $p_{\g}^{(\E)}$ can be attained by a different optimal measurement. This shows that an optimal measurement is protected from a noisy channel. It is also found that the guessing probability is enhanced, i.e., $p_{\g}^{(\mathcal{\widetilde{S}}\E) } > p_{\g}^{( \E)} $ for all $p \in [0,1/2]$. From a cryptographic point of view, the higher the distinguishability the lesser the error, meaning that a QBER can be made lower by which the security condition may be improved.

We now derive the security condition for MP QKD protocols. The security analysis of P\&M QKD protocols can be equivalently performed in the entanglement-based picture. Thereby, as it has been explained, two honest parties share a Bell-diagonal state on a single-copy level \cite{kraus2005, renner2005a, renner2008}, 
\bea
\rho_{AB}^{(\N_{\bf p})} = p_0 |\phi_1\rangle \langle \phi_1| + p_z |\phi_2 \rangle \langle \phi_2 | + p_x |\phi_3 \rangle \langle \phi_3 | + p_y |\phi_4 \rangle \langle \phi_4 |  \nonumber
\eea
where $|\phi_2 \rangle = (|00\rangle - |11\rangle )/\sqrt{2}$, $|\phi_3 \rangle = (|01\rangle + |10\rangle )/\sqrt{2}$, and $|\phi_4 \rangle = (|01\rangle - |10\rangle )/\sqrt{2}$. From the shared state, an error rate denoted by $Q^{( \N_{\bf p})}$ in standard P\&M QKD protocols can be computed, 
\bea 
Q^{(\N_{\bf p})} = \tr[\rho_{AB}^{(\N_{\bf p})} (|01\rangle \langle 01| + |10\rangle \langle 10| )] =p_x + p_y. \label{eq:qn}
\eea
The BB84 protocol introduces a channel such that measurements in the $x$- and $z$-basis are symmetric \cite{ bb84}. By this constraint, the parameters can be written as follows
\bea
p_0 = 1-2Q^{(\N_{\bf p})}  +x,~ p_x = p_z = Q^{(\N_{\bf p})}-x,~ p_y =x ~~~~~\label{eq:bb84p}
\eea
where a free parameter $x$ is chosen to minimize the secret key rate. Note that the channel is supposed to be under an eavesdropper's control \cite{christandl2004}. The parameter also depends on the secret key agreement protocol that two honest parties apply. It has been found that $x=(Q^{(\N_{\bf p}) } )^2$ when two honest parties apply one-way information reconciliation and privacy amplification \cite{christandl2004}, and $x=0$ when the advantage distillation with two-way communication is incorporated before the one-way protocol \cite{bae2007, branciard2005}. For the six-state protocol, we have $p_0 = 1-3Q^{(\N_{\bf p})}  /2$ and $p_x = p_y = p_z = Q^{(\N_{\bf p})}  /2$ \cite{christandl2004}. This also introduces a depolarization channel which therefore contains the measurement protection. 

With the measurement protection, two honest parties are connected by a Pauli channel. In the entanglement-based scheme, the shared state is characterized as
\bea
\rho_{AB}^{ (\mathcal{\widetilde{S}}\N_{\bf p}) } = (1-\eta)  |\phi_1\rangle \langle \phi_1| +  \frac{\eta}{4} \mathbbm{I},~\mathrm{with}~\eta = \frac{4(1-p_0)}{3}. ~~~~\label{eq:sqtn}
\eea
The error rate is also transformed accordingly as follows,
\bea 
Q^{(\N_{\bf p})} ~\mapsto ~Q^{ (\mathcal{\widetilde{S}}\N_{\bf p}) } = \frac{2}{3} (Q^{(\N_{\bf p})}  +p_z). ~~\label{eq:qtn}
\eea
From Eqs. (\ref{eq:qn}) and (\ref{eq:qtn}), QBERs can be made lower, i.e., $Q^{ (\mathcal{\widetilde{S}}\N_{\bf p})} < Q^{(\N_{\bf p})}$, by the measurement protection for channels satisfying $p_0 < 1-3p_z$. For instance, for channels with $p_z =0$ the measurement protection can always make the error-rate lower.

After distributing quantum states, two honest parties can apply one-way communication protocols to distill a secret key. Against general attacks, the critical values of QBERs have been found as $11\%$ in the BB84 protocol \cite{shor2000} and $12.7\%$ in the six-state protocol \cite{bruss2002}. These can be improved by the pre-processing in the preparation of states, up to $12.4\%$ and $14.1\%$ in the BB84 and six-state protocols, respectively \cite{kraus2005, renner2005a, renner2008}. Thus, an MP QKD protocol can tolerate QBERs up to $Q^{ (\mathcal{\widetilde{S}}\N_{\bf p}) } < 12.7\%$ and $Q^{ (\mathcal{\widetilde{S}}\N_{\bf p}) } <14.1\%$ without and with the classical pre-processing, respectively. In Eq. (\ref{eq:qtn}), it is shown how QBERs are also manipulated by the measurement protection. For the MP BB84 protocol with one-way secret key distillation, we have the relation as follows, 
\bea
Q^{ ( \N_{\bf p}) } = 1- \sqrt{ 1- 3 Q^{ (\mathcal{\widetilde{S}}\N_{\bf p}) }  /2} \label{eq:rel}
\eea
with the optimal parameter $x=(Q^{(\N_{\bf p}) } )^2$. Then, the MP BB84 protocol can tolerate error rates up to $Q^{(\N_{\bf p}) } = 10\%$ and $Q^{(\N_{\bf p}) } = 11.2\%$ without and with the classical pre-processing, respectively. It is shown that the cost to pay for protecting a measurement without the verification of a channel is to have a little lower key rate and subsequently slightly lower values of critical QBERs for QKD protocols.

The security conditions can be improved in the key agreement scenario with two-way communication of the advantage distillation \cite{maurer1999}, followed by one-way communication of information reconciliation and privacy amplification. Higher error-rates can be tolerated by post-selection on sifted bits by applying a two-way communication protocol beforehand. The advantage distillation works as follows. Given the sifted bits $(m_{i}^{(A)}, m_{i}^{(B)} )_{i=1}^n$, the advantage distillation begins by Alice's generating a secret bit $s_A$ and chooses a block of $k$ bits. She computes $a_i = s_A \oplus m_{i}^{(A)} $ for the blocks and announces the resulting values $(a_i)_{i=1}^m$  to Bob publicly, who then computes $b_i = a_i \oplus m_{i}^{(B)}$ on his side. Bob accepts the $k$-block $(b_i)_{i=1}^m$ only when the block contains the same values, i.e., when $b_1=\cdots=b_k$. Finally, Bob announces to Alice publicly if he accepts or discards the block. For the post-selected bits, two honest parties proceed to one-way information reconciliation and privacy amplification. 

Against general attacks, the security condition has been obtained \cite{bae2007, branciard2005} : the secret key rate is positive, i.e., there exists a block size $k$ such that secret bits can be distilled from a shared channel $\N_{\bf p}$, by the aforementioned key agreement protocol whenever the channel satisfies the condition $(p_0 - p_z)^2 > ( p_0 + p_z) ( p_x + p_y)$. Note also that the security condition can be obtained by the undistillability constraint that symmetric extensions of quantum states do not allow a positive key rate \cite{myhr2009}. Applying the condition to MP QKD protocols, we find that from Eqs. (\ref{eq:sqtn}) ad (\ref{eq:qtn}), secret key can be distilled whenever the following is satisfied,
\bea
p_0 > \frac{5 + 3\sqrt{5}}{20},~\mathrm{or~equivalently}~ Q^{ (\mathcal{\widetilde{S}}\N_{\bf p})}  < 27.6\%. \label{eq:sec}
\eea
Note that the shared state in Eq. (\ref{eq:sqtn}) is entangled for $Q^{ (\mathcal{\widetilde{S}}\N_{\bf p})}<33.3\%$, in which there exists a measurement of Alice and Bob such that secret correlations exist in a joint distribution \cite{acin2005}. For the BB84 protocol with two-way key distillation, we recall that the optimal parameter is given by $x=0$ in Eq. (\ref{eq:bb84p})  \cite{bae2007, branciard2005}. By the measurement protection, the QBER is transformed as, $Q^{ (\mathcal{\widetilde{S}}\N_{\bf p}) } = 4 Q^{ (\N_{\bf p}) }  /3$. From the condition in Eq. (\ref{eq:sec}), it is straightforward to see that secret key can be distilled whenever $Q^{(\N_{\bf p})}<20.7\%$, that shows an improvement over the existing bound $20\%$ \cite{bae2007}. Note that the shared state is entangled for $Q^{(\N_{\bf p}) }<25\%$ in which two honest parties can still share secret correlations by a measurement \cite{acin2005}.

\begin{figure}[t]
\begin{center}
\includegraphics[width=3.4in]{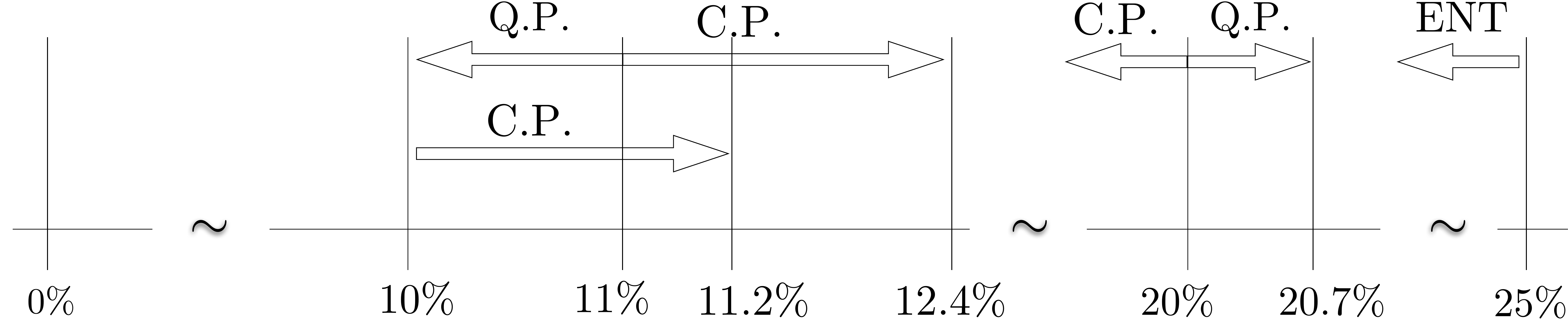}
\caption{  Security bounds in the BB84 protocol are compared. In the standard BB84 protocol with one-way key distillation, secret key can be distilled for $Q^{ (\N_{\bf p})} <11\%$. By the classical pre-processing (C.P.), it is improved up to $12.4\%$. After the measurement protection with quantum pre- and post-processing (Q.P.), the bounds are $10\%$ and $11.2\%$ without and with the classical pre-processing. When the advantage distillation protocol is incorporated, the bound is given by $20\%$ that is improved to $20.7\%$ by quantum pre- and post-processing. The upper bound is $25\%$ which is from the entanglement condition (ENT). No secret key can be distilled from separable states. }
\label{bound}
\end{center}
\end{figure}

In the measurement protection, the quantum pre- and post-processing by the application of unitaries play a key role.  Although one may find that random applications of the unitaries cause addition of noise to quantum states, it turns out that they can in fact improve distinguishability of states and the security of P\&M QKD protocols. This shares some similarities to the classical pre-processing, which also adds classical noise with some probability but has improved the one-way secret key rate \cite{kraus2005, renner2008}. In the BB84 protocol, the classical pre-processing improves the critical QBERs up to $12.4\%$ \cite{kraus2005, renner2005a, renner2008} over the previous one $11\%$ \cite{shor2000} . When two-way communication is included in the key distillation protocol, the security bound is found to be $20\%$ \cite{bae2007, PhysRevA.75.012316}, which is then improved to $20.7\%$ by the quantum pre- and post-processing. It is found that the quantum pre- and post-processing makes two honest parties symmetric in the distribution of quantum states, see Eq. (\ref{eq:dep}). In this way, a QBER of asymmetric channels may be made even lower so that it can be tolerated in secret key distillation. In particular, this turns out to be well-fitted to the two-way distillation protocol where two parties are symmetric after the advantage distillation. The classical pre-processing makes a channel of two honest parties even more asymmetric. In this way, two honest parties may have a higher error rate in general but an eavesdropper may get even less information, so that the one-way key rate is improved and higher values of a QBER are tolerated. In addition, we have found that the quantum pre- and post-processing does not improve the one-way secret key rate, e.g. $10\%$ in the MP BB84 protocol. Note also that the classical pre-processing fails to improve the two-way secret key rate \cite{bae2007}. 

In conclusion, we have presented MP QKD protocols that deal with uncontrollable unknown sources of noise in the alignment of a measurement setting of a receiver when quantum states are distributed over a long distance. The advantages are twofold. Firstly, the verification of a channel during the transmission is in fact circumvented. This saves a significant amount of experiment resources and makes QKD protocols more feasible in real-world applications. An optimal measurement prepared for an ensemble of states would be ever optimal even if the states are sent over a noisy channel. Secondly, the measurement protection can improve the security condition of QKD protocols. The cost to realize the measurement protection amounts to implementation of a few local unitaries before and after the transmission. All of our results are supported by experimental demonstrations: QBERs can be made lower and quantum states can be better discriminated by the measurement protection. 


Our results show that the scheme of measurement protection makes P\&M QKD protocols more feasible in real-world applications. Importantly, a noisy channel does not have to be verified at all for optimal detection of noisy and unknown resulting states. It suffices to estimate the error rate only. As a byproduct, distinguishability can be improved so that quantum communication can be established at an even longer distance, and also an error rate can be made even lower such that a QKD protocol is more resilient to noise. To improve the level of security further, it would be interesting to investigate what assumptions can be relaxed in MP QKD protocols, similarly to device-independent \cite{acin2007} or measurement-device-independent \cite{lo2012} scenarios. Our results have shown that the BB84 protocol can tolerate an error rate of up to $20.7\%$. Note, however, that the upper bound is $25\%$ over which no secret key can be distilled. It remains an open question whether quantum correlations in the gap can be used to distill a secret key or not. 

 \section*{Acknowledgements}


S.K. and J.B. are supported by National Research Foundation of Korea (2019M3E4A1080001), an Institute of Information and Communications Technology Promotion (IITP) grant funded by the Korean government (MSIP), (Grant No. 2019-0-00831, EQGIS), and ITRC Program(IITP-2019-2018-0-01402). H.K, Y.-H.K, J.-K.K, B.-S.C., and C.J.Y. are supported by the ICT R\&D program of MSIP/IITP (Grant No. 1711081002, The technology development of transmitter and receiver integrated module in a polarization-based free-space quantum key distribution for short-range low-speed moving quantum communication). C.M.K. is supported by Iniziativa Specifica INFN-DynSysMath.


\bibliography{odqv6_biblio.bib}
\bibliographystyle{apsrev4-1}


\end{document}